\pdfoutput=1 
\documentclass{natureprintstyle}

\usepackage[utf8]{inputenc} 
\usepackage[left]{lineno} 
\usepackage{cite} 
\usepackage{graphicx} 
\usepackage{url} 
\usepackage{amsmath}	
\usepackage{amssymb}	
\usepackage[export]{adjustbox} 
\usepackage[normalem]{ulem} 

\def\kms{km s$^{-1}$}

\usepackage[T1]{fontenc}
\usepackage{relsize}

\def\fdg{.\!\!^\circ}

\usepackage{tikz}

\usepackage[symbol]{footmisc}

\newcounter{firstbib}

\newif\iffinal
\iffinal\renewcommand{\includegraphics}[2][]{}\fi

\newcommand\aap{{Astron. Astrophys.}}
\newcommand\apjl{{Astrophys. J. Lett.}}
\newcommand\apj{{Astrophys. J.}}

\newcommand\aj{{Astron. J.}}
\newcommand\araa{{Ann. Rev. Astron. Astrophys.}}

\newcommand\nat{{Nature}}

\newcommand\mnras{{Mon. Not. R. Astron. Soc.}}

\newcommand\pasj{{Publ. Astron. Soc. Japan}}

\title{A Keplerian disk with a four-arm spiral birthing an episodically accreting high-mass protostar}

\author{R. A. Burns,$^{1,2,3 \star}$ 
Y. Uno,$^{4}$ 
N. Sakai,$^{3,5}$
J. Blanchard,$^{6}$
Z. Rosli,$^{7}$
G. Orosz,$^{8}$
Y. Yonekura,$^{9}$
Y. Tanabe,$^{9}$
K. Sugiyama,$^{5}$
T. Hirota,$^{1,10}$
Kee-Tae Kim,$^{3,11}$
A. Aberfelds,$^{12}$
A. E. Volvach,$^{13}$
A. Bartkiewicz,$^{14}$
A. Caratti o Garatti,$^{15}$
A. M. Sobolev,$^{16}$
B. Stecklum,$^{17}$
C. Brogan,$^{18}$
C. Phillips$^{19}$
D. A. Ladeyschikov,$^{16}$
D. Johnstone,$^{20}$ 
G. Surcis,$^{21}$
G. C. MacLeod,$^{22,23}$
H. Linz,$^{24}$
J. O. Chibueze,$^{25,26}$
J. Brand,$^{27}$ 
J. Eisl\"offel,$^{17}$
L. Hyland,$^{28}$ 
L. Uscanga,$^{29}$
M. Olech,$^{30}$
M. Durjasz,$^{14}$
O. Bayandina,$^{31}$
S. Breen,$^{32}$
S. P. Ellingsen,$^{28}$
S. P. van den Heever,$^{23}$
T. R. Hunter,$^{18}$
X. Chen,$^{33,34}$}

\begin{document}
\maketitle

\begin{affiliations}
[Affiliations list moved to end of document]\\
 $^{\star}$ e-mail: ross.burns@nao.ac.jp
\end{affiliations}





\begin{abstract}
High-mass protostars (M$_{\star} >$ 8 M$_{\odot}$) are thought to gain the majority of their mass via short, intense bursts of growth. This episodic accretion is thought to be facilitated by gravitationally unstable and subsequently inhomogeneous accretion disks. Limitations of observational capabilities, paired with a lack of observed accretion burst events has withheld affirmative confirmation of the association between disk accretion, instability and the accretion burst phenomenon in high-mass protostars.
Following its 2019 accretion burst, a heat-wave driven by a burst of radiation propagated outward from the high-mass protostar G358.93-0.03-MM1.
Six VLBI (very long baseline interferometry) observations of the raditively pumped 6.7 GHz methanol maser were conducted during this period, tracing ever increasing disk radii as the heat-wave propagated outward. Concatenating the VLBI maps provided a sparsely sampled, milliarcsecond view of the spatio-kinematics of the accretion disk covering a physical range of $\sim$ 50 - 900 AU. We term this observational approach `heat-wave mapping'.
We report the discovery of a Keplerian accretion disk with a spatially resolved four-arm spiral pattern around G358.93-0.03-MM1. This result positively implicates disk accretion and spiral arm instabilities into the episodic accretion high-mass star formation paradigm. 

\end{abstract}


Accretion bursts have now been observationally confirmed in three high-mass protostars, S255IR-SMA1\cite{Garatti17}, NGC6334I-MM1\cite{Hunter17}, and G358.93-0.03-MM1\cite{Stecklum21a}, where accompanying luminosity enhancements, indicative of increased accretion rates were recorded at infrared wavelengths, lasting as long as several years (as for S255IR-SMA1\cite{Szymczak18} and NGC 6334I-MM1\cite{Macleod18}) to as short as several months (as for G358.93-0.03-MM1\cite{Volvach20a,Macleod19a}). The apparently analogous FU-Orionis and EX-Lupi accretion bursts of low mass protostars [ref,\cite{Hartmann96}] have long been postulated to arise as the result of non-continuous, episodically varying accretion rates provided to the protostar from a gravitationally unstable, and thereby inhomogenius, accretion disk\cite{VorobyovBasu05}. Simulations specific to the high-mass star formation regime have shown success in replicating the observed accretion bursts, which are explained as the episodic assimilation of disk clumps and spiral arm segments that migrate through the accretion disk toward the protostar as angular momentum is dissipated via viscosity and the induced torque of disk gravitational instability (GI)\cite{Meyer17,Meyer21a}. 
However, the relative rarity of high-mass protostars, in addition to their embedded nature, large distances and short time spent in the burst phase makes it extremely challenging to obtain the high spatial resolution data needed to observationally confirm the association of GI-induced accretion disk instabilities\cite{Jankovic19} and the accretion burst phenomenon in high-mass protostars.

Maser emission has recently demonstrated proficiency in identifying accretion bursts, with all three high-mass protostellar accretion burst events to date exhibiting maser flares\cite{Fujisawa15,Brogan18,Moscadelli17,Szymczak17a,Gordon18,Macleod19a}.  
The high brightness temperatures, narrow spectral profiles and compact emission of masers make them suitable for VLBI (very long baseline interferometry) observations, which in turn provide milliarcsecond resolution spatio-kinematic diagnostics of structures associated with accretion bursts, unhindered by spectral line broadening effects affecting thermal gas kinematic investigations in protostellar disks [ex. ref\cite{Ginsburg18}].

\null 

\subsection{High-mass protostar G358.93-0.03 MM1}

Of the three known cases, G358.93-0.03-MM1 (hereafter ``G358-MM1'') is the most recent and most intensely studied accretion burst event of a high-mass protostar. Acquisition of observational data recording the burst event was aided by a coordinated survey and follow-up campaign conducted by the Maser Monitoring Organisation (M2O; https://www.masermonitoring.org), which is a community of multi-wavelength observers, astrophysicists and maser theoreticians pursuing research into time-variable astrophysical phenomena traced by maser emission.

The G358.93-0.03 high-mass star forming region was first identified by its 6.7 GHz methanol maser emission\cite{Caswell10a} which was consistently $<10$ Jy on the occasions it was observed\cite{Caswell10a,Chambers14,Rickert19,Hu16a}. New investigations into G358-MM1 were initiated in response to a flare of the methanol $5_1 \xrightarrow{} 6_0~A^+$ maser transition at 6.7 GHz, reported by the iMet maser monitoring programme lead by Ibaraki University using the Hitachi 32m radio telescope\cite{Yonekura16}. This maser transition is excited by 20-30 $\mu m$ far infrared radiation from 100-150 K dust\cite{Cragg05}. The temperatures and densities required to produce maser emission at this transition limit its detection exclusively to regions of high-mass star formation\cite{Breen10}. 

Tracking the evolution of the maser emission effectively traces the evolution of dust heating in the protostellar disk. The 6.7 GHz methanol maser flare in G358.93-0.03 began on the 14$^{th}$ Jan 2019 \cite{Sugiyama19}, reached peak intensity of 900 Jy in March followed by a gradual reduction of activity until June when the flare was determined to have subsided thereafter, concluding the $\sim$5 months, or $\sim$150 days, of main flare activity. In the wake of the flare the maser remained at a stable but slowly decreasing $\sim$20 Jy, seemingly returning to its pre-flare state. 

The M2O-led investigation of G358.93-0.03 revealed it to be a cluster of 7 millimeter cores, of which G358-MM1 was the progenitor of the accretion burst and associated maser flare\cite{Brogan19a}. At the time of writing 27 new maser transitions have been discovered toward G358-MM1, and several rarely seen maser transitions were reported to have flared in association with the accretion burst event\cite{Breen19a,Brogan19a,Macleod19a,Chen20a,Chen20b} indicating a peculiar and short-lived physical and radiative environment. 
The 2019 accretion burst was investigated using radiative transfer modelling based on the pre-, mid- and post-burst spectral energy distribution (SED) whereby it was determined that G358-MM1 gained $5.3^{+11.1}_{-3.3}\times 10^{-4} M_{\odot}$ of mass at a rate of $\dot M _{acc} = 3.2^{+5.4}_{-3.0} \times 10^{-3} M_{\odot} ~ \rm yr^{-1}$ [ref, \cite{Stecklum21a}]. Independent agreement was obtained for estimates of the stellar mass ($\sim 9 ~\rm{to} ~15~ M_{\odot}$) and system inclination ($\sim 10^{\circ} ~\rm{to}~ 35^{\circ}$ from face-on) in both analyses of dynamics traced by spectral line data\cite{Chen20b,Brogan19a} and the findings of infrared radiative transfer modelling\cite{Stecklum21a}.

\begin{table*}[h!]
\centering
\caption{Observation parameters: \label{TAB1}}
\small
\begin{tabular}{llllcclc}
\hline
Facility& Participating &Observation& Observation & Spect. res.  & Image rms  & Beam$^{\dagger}$ & Ast. uncertainty$^{r}$\\ 
name  & stations  & date       & Code      & [kHz]     & [mJy/beam/channel]   & [mas x mas / degrees] & [mas]\\ 
\hline
  LBA  &  AT CD HH HO MP WA  & 2 Feb 2019  & VC026A  & 0.98 & 77.5 &  $10.1\times 3.0$ / -51.43 & 1.30\\
  LBA  &  AT CD HO MP PA WA  & 28 Feb 2019 & VC026C  & 0.98 & 231.7 & $8.6\times 3.6$ / -89.69 & 1.22\\
  EVN  &  JB EF MC TR YS SR HH WB   & 13 Mar 2019 & RB005   & 1.95 & 284.7 & $37.4 \times 11.6$ / -72.44 & 5.03\\
  VLBA &  BR FD HN KP LA MK NL OV PT & 19 May 2019 & BB414   & 3.91 & 40.4 & $17.2\times 2.6$ / -18.65 & 1.99\\
  VLBA &  BR FD HN KP LA MK NL OV PT  & 7 Jun 2019  & BB412   & 3.91 & 88.7 &  $17.5\times 8.4$  / 3.89& 2.60\\
  LBA  &  AT CD HH HO KE MP PA WA   & 28 Sep 2019 & V581A   & 0.98 & 42.9 & $37.4\times 11.6$ / -72.44& 1.55\\
\hline 
\end{tabular} 
\begin{flushleft}
$^{\dagger}$ Synthesized beam semi major, (x) semi minor axes and (/) position angle in degrees\\
$^{r}$ Astrometric uncertainties for each VLBI epoch were derived using the estimation method described in (ref,\cite{Asaki07}). Individual spot positional uncertainties were calculated by this number, and the 2D Gaussian fitting error for each spot, and the reference quasar, added in quadrature.
\end{flushleft}
\end{table*}

Thermal and maser emission from centimeter to millimeter wavelengths presented continuous curvelinear structures in the position-velocity diagrams of subarcsecond resolution observations made with the Atacama Large Millimeter Array (ALMA)
\cite{Brogan19a} and the the Very Large Array (VLA)\cite{Chen20b,Bayandina22}. However previous observations did not have the angular resolution necessary to resolve such structures, which were noted to vary depending on the choice or combination of molecular line tracers under consideration\cite{Chen20b,Brogan19a}. The presence of unresolved velocity structure indicated the need for re-investigation with more complete, higher angular resolution data.

Two VLBI observations of the 6.7 GHz methanol maser transition at day 19 and 45 since the flare onset (14$^{th}$ Jan 2019) have already been reported, revealing a `heat-wave' of accretion energy propagating outward from G358-MM1 at subluminal speeds\cite{Burns20b}. Here we show the VLBI monitoring results of six VLBI observations of G358-MM1 on days 19, 45, 58, 125, 144, 257 and present new analyses and more detailed and conclusive investigation into its disk kinematics and substructure.

\subsection{VLBI maps of 6.7 GHz methanol masers in G358-MM1}

VLBI observations of G358-MM1 conducted using the Long Baseline Array (LBA), the European VLBI Network (EVN) and the Very Long Baseline Array (VLBA) were acquired via Target of Opportunity observation requests. Details of the observation setup, array characteristics, data reduction, analyses and data availability are given in the Methods section.
Since the varying baseline lengths and orientations of VLBI arrays participating in this campaign yield epoch-specific synthesized beam shapes and sizes we omit detailed discussion of the evolution of spectral profiles and individual maser feature sizes and instead focus on spatio-kinematics at the milliarcsecond scale.

\begin{figure*}[ht!]
\begin{center}
  \includegraphics[width=0.99\textwidth]{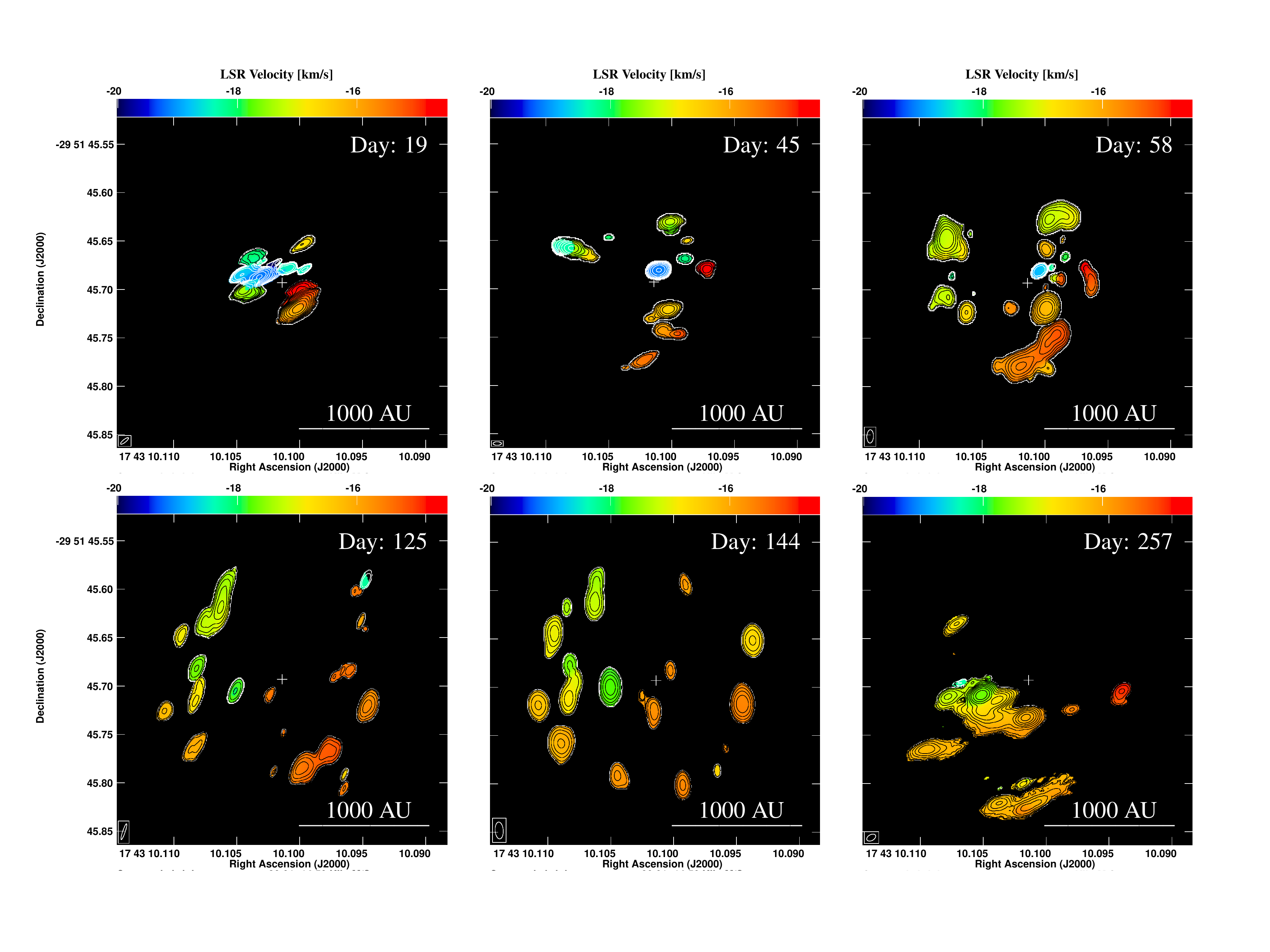}
\end{center}
\caption{Moment maps of the 6.7 GHz methanol masers in G358-MM1. 1st moment (velocity) maps are shown in colour, where the source velocity is $v_{\rm LSR}=-16.5 \pm 0.3$ km s$^{-1}$ [ref\cite{Brogan19a}]. Contours show the zeroth moment (flux density) map for each epoch, produced for emission above a 5 times the rms listed in Table~\ref{TAB1}. Contours begin at 1 Jy beam$^{-1}$ m s$^{-1}$ Coordinates and colour scales for velocities are common to all maps. The white cross indicates the position of the millimeter core\cite{Brogan19a}. The time of the observation in days since the flare onset is shown in the top right of each panel. \label{Fig1}} 
\end{figure*}

\begin{figure*}[ht!]
\begin{center}
\includegraphics[width=0.92\textwidth]{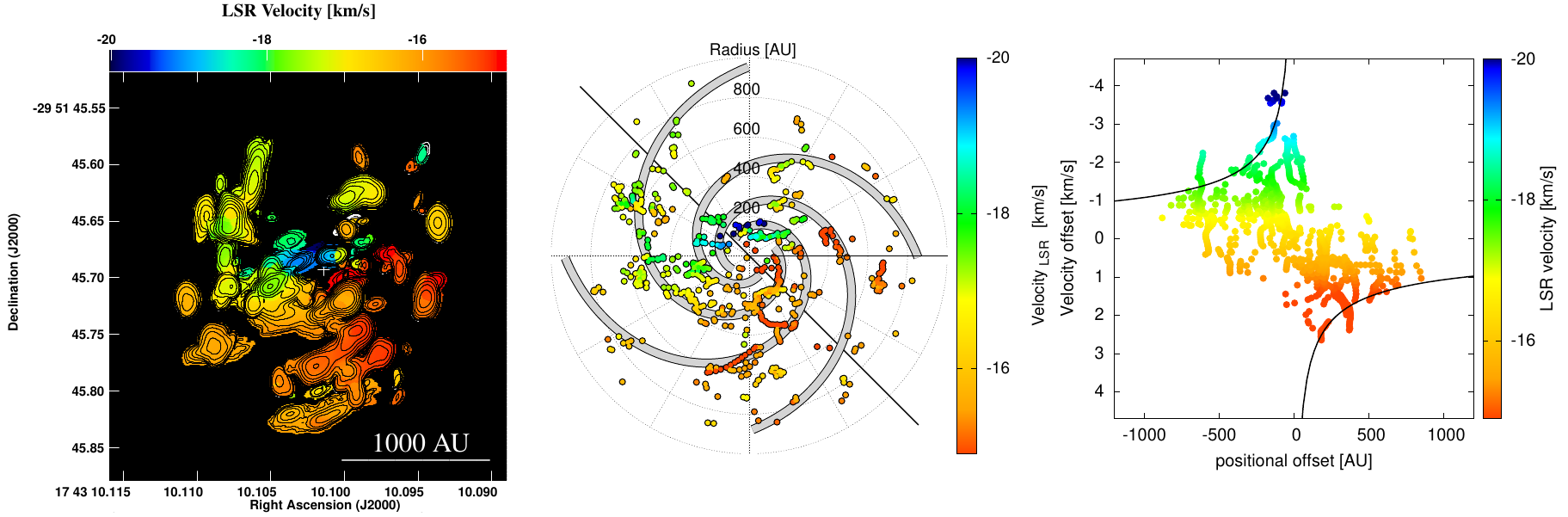} 
\end{center}
\caption{ {\bf Methanol maser spatio-kinematics in G358-MM1.} (\emph{Left}) Zeroth (contours) and first (colours) moment maps of the 6.7 GHz methanol maser emission in G358-MM1 created by combining all six VLBI maps shown in Figure~\ref{Fig1}, in which Zeroth moments were summed and first moments were averaged. Contours begin at 1 Jy beam$^{-1}$ m s$^{-1}$ and increase in a geometric sequence (1,2,4,8,16,...). The white cross indicates the position of the G358-MM1 millimeter core\cite{Brogan19a}. (\emph{Middle}) shows the spotmap of the combined six epoch data sets centered on the G358-MM1 position. The black line indicates the direction of largest velocity gradient to which the position-velocity cut was taken. Spiral arms identified in this work (see Methods) are plotted as thick grey lines. (\emph{Right}) shows  the position velocity diagram of maser spots taken along the cut indicated in the central panel. The Keplerian function for a 11.5 M$_{\odot}$ enclosed mass (see Methods) is shown as a black line. \label{Fig2}}
\end{figure*}


VLBI maps of the flux density and velocity of the 6.7 GHz methanol maser emission shown in Figure~\ref{Fig1} reveal concentric ring-like distributions of maser features centered on the high-mass protostar G358-MM1, in which the extent of the maser ring increases dramatically throughout the campaign. In the final epoch the ring-like distribution gave way to a more centrally compact, orderless distribution. The six individual snapshots shown in figure exhibit a slight enhancement in maser brightness at directions north-east and south-west of the protostar's location. This can be understood in the context of an inclined disk, in which gas with the longest velocity coherent path lengths toward the observer appear in these regions. Methanol maser emission was typically within $\pm 5$ \kms of the systemic line of sight velocity of the target of $16.5 \pm 0.3$ \kms determined by ALMA observations of dense core gas and disk gas tracers\cite{Brogan19a}. A velocity gradient in which blue shifted emission is seen to the north-east, and redshifted emission to the south-west of G358-MM1 is evident in the maser emission maps and is present at all observing sessions indicating that all epochs trace a common kinematic structure.

Maser translocation observed in G358-MM1 was $> 1$\% $c$; significantly faster than the typical gas motions associated with disk rotation and shock shell expansion motions. Furthermore maser emission cannot be maintained when bulk motions exceeds 10 km s$^{-1}$[ref\cite{Garay02a}], indicating that the observed redistribution of maser emitting regions was due to the radially directed propagation of maser excitation conditions (heat) rather than physical gas motions. The speed of the relocation of methanol masers in G358-MM1 has been discussed in the context of radiative transfer [see Section 7 of ref\cite{Stecklum21a}] based on the two VLBI epochs published in [ref\cite{Burns20b}] . An expanded radiative transfer modelling employing the findings of the six VLBI epochs presented here is now underway (Stecklum et al. in prep.).

Maser emission was excited at ever increasing radial distances from the protostar as the accretion-driven heat-wave propagated outwards. The six VLBI epochs effectively sample six radially incremented static snapshots of the spatio-kinematics of gas in the region around the protostar. Combining the six concentric snapshots produces a single milliarcsecond-scale, discretely sampled map of the kinematic structure traced by maser-emitting gas around G358-MM1 (Figure~\ref{Fig2}, \emph{left and middle}). We term this technique of combining VLBI radial snapshots to produce a single map as `heat-wave mapping'.

The outermost methanol maser detected in our observations was identified in the penultimate epoch at $r_{max} = 135.98 \pm 2.60$ milliarcseconds, corresponding to a physical radial distance of $917.87 \pm 17.55$ AU. The uncertainty is derived from the astrometric error of the maser spot coordinates and its fitting error. Theis radius matches closely to the $\sim$940 AU disk scale determined from millimeter emission imaged by ALMA\cite{Brogan19a}. Beyond this radial distance the absence of 20-30 $\mu m$ emission needed to pump methanol masers would force the truncation of the maser-traced disk to the same radius as the dust disk. The $\sim 900$ AU radius of the disk is consistent with the expected 500 - 1500 AU size range deemed appropriate for the high-mass regime in modelling\cite{Krumholz07} and is consistent with observations of disks around high-mass protostars\cite{Johnston20,Bik04,Kraus10,Wheelwright10,Boley13,Ilee13,Tannus17} (see also ref,\cite{Beltran16} and references therein). 

Analysis of the position-velocity profile of the combined data set revealed that the 6.7 GHz methanol masers in G358-MM1 were excited in a disk of gas exhibiting Keplerian rotation (Figure~\ref{Fig3}, \emph{left}). Conversely, rotating envelopes and torii exhibit velocity profiles with a power law slope index of -1, transitioning to a shallower slope consistent with -0.5 for Keplerian rotation inside the centrifugal barrier\cite{Ohashi14,Aso15}. The Keplerian rotation profile of $v \propto r^{-0.510 \pm 0.095}$ in G358-MM1 shows no transition to a steeper index even at the largest radii traced by masers, indicating that the 6.7 GHz methanol masers trace disk radii within the centrifugal barrier.

The disk inclination of was determined by analysing the ellipticity of the maser ring distribution in the fifth VLBI epoch, chosen for its large radius and completeness (See Figure~\ref{Fig3}, \emph{right}). An inclination estimate of $21.4 \pm 4.7 ^{\circ}$ was obtained. The enclosed dynamic mass derived from the position-velocity profile and uncoupled from the observational effects of disk inclination, was estimated to be $M = 11.5 \pm 4.8$ M$_{\odot}$ (see Methods). Both parameters are consistent with previous works\cite{Chen20b,Stecklum21a}. Disk parameters were determined assuming the commonly adopted $D=6.75^{+0.37}_{-0.68}$ kpc distance estimate to G358-MM1 derived using the Bayesian statistics Galactic distance estimation tool provided by the BeSSeL project\cite{Reid14}.

\subsection{Spiral structure in G358-MM1}

\null 

Methanol masers in the G358-MM1 disk are distributed in spiral arm structures with a logarithmic relationship between disk radius, $R$, and azimuth angle, $\phi$. Spiral arm morphology was determined by Monte Carlo Markov Chain fitting to maser data in $\phi-ln(R)$ space, identifying two spiral arms at a $\sim \pi/2$ azimuth angle separation (See Figure~\ref{Fig4}). The spiral pattern has a pitch angle of $21.2 \pm 0.2$ degrees and pervades the full between $50 \sim 900$ AU. Further symmetric pair spiral arms were identified using two-dimentional cross-correlation of a spiral function and the full maser data set. The analyses, detailed in the Methods section, present an accretion disk with 4 spiral arms with 90$^{\circ}$ rotational symmetry.

The majority of 6.7 GHz methanol masers are collocated with one of the four identified spiral arms, however a small contribution of maser emission also appears in the inter-arm regions (See Figure~\ref{Fig5} \emph{upper panel}). The detectability of maser emission is influenced by gas column density. Thus, since the disk is near to a face-on inclination, the inter-arm maser emission may arise from gas condensations above the disk-plane or may represent additional minor condensations or minor sub-structures within the disk. However, aside from the spiral arm structures, a Random Sample Consensus (RANSAC) analysis of the distribution of masers did not discern any identifiable logarithm structures or radial structures, indicating that the spiral arms are the dominant features traced by 6.7 GHz methanol masers in the heat-wave mapping of G358-MM1.

 Various spiral morphologies can arise in protostellar disks, each depending on the underlying process which instigated the instability.
 Fly-by encounters induce spiral patterns initially dominated by a single arm with shocklike structure, with a second arm forming later as a result of an induced offset between the protostar and the gravitational center of the disk\cite{Pfalzner03}. These arms ($m=1$ or $m=2$, where $m$ denotes the number of arms) have pitch angles that diverge from a purely logarithmic morphology and depart from the disk mid-plane, owing to perturbation-induced disk warping\cite{Cuello19,Cuello20,Pfalzner03}. These spatio-kinematic characteristics are not apparent in the disk of G358-MM1 as systematic super- or sub-keplerian velocity residuals were absent from the maser data (See Figure~\ref{Fig6}). The morphological characteristics of spiral patterns in protostellar disks induced by companions and self-gravity were compared in a work by [ref\cite{Forgan18}]; a spiral pattern with logarithmic arms of constant pitch angle, complete disk coverage and 90$^{\circ}$ rotational symmetry (i.e. $m=4$) as seen in the masers of G358-MM1, indicates an origin in GI rather than the tidal influence of a companion object or accretion streams. 
 
 Regarding the possible presence of close companions in the inner disk, this is somewhat constrained. Since the 6.7 GHz methanol maser transition is only excited in the presence of high-mass protostars\cite{Breen10}, the G358-MM1 disk must contain at least one high-mass ($>8$ M$_{\odot}$) protostar in its center. Given that the estimated stellar mass of G358-MM1 is $\sim 10 - 12$ M$_{\odot}$ (this work; [refs\cite{Chen20b,Stecklum21a}]) any close companion yet undetected in the inner disk of G358-MM1 could be only a few solar masses at most. Thus the central mass budget is likely dominated by one high-mass protostar.
 
 \subsection{Spiral instabilities invoked by self-gravity}
 
 \null

Spiral pattern gravitational instabilities trap dust into the spiral arms\cite{Dipierro15} and lead to enhanced gas surface densities in the arms\cite{Ahmadi19}. Accretion bursts in such systems produce warm dust temperature distributions that are subsequently more pronounced in the spiral arm regions\cite{Bae14}. It is therefore plausible that pre-existing spiral arms in G358-MM1, with enhanced dust and gas column densities, provided the 20-30 $\mu m$ radiation and maser gas path length needed to excite observable 6.7 GHz maser emission in the presence of the heat-wave. Subsequently, selective production of methanol masers in these regions and less maser emission in the lower density inter-arm regions can explain the detection of the spiral structure in G358-MM1. A similar isolation of methanol maser emission was seen in the dusty streams of the infalling rotating envelope of another face-on high-mass protostar G353.273+0.641\cite{Motogi17}.


The morphologies of GI-induced spiral instabilities relate to the degree of instability, evaluated by the Toomre Q parameter\cite{Toomre64}, which can be approximated as:
\begin{eqnarray}
Q ={c_s \kappa  \over \pi G \Sigma} = f {M_\star  \over M_d} {H \over r}
\label{eq:Q}
\end{eqnarray}

\noindent where $c_s$ is the sound speed, $\kappa$ is the epicyclic frequency of disk rotation and $\Sigma$ is the disk surface density. The equation can be approximated to the rightmost expression in which $M_\star$ is the protostellar mass, $M_d$ is the disk mass, $H$ and $r$ are the disk scale hight and radius respectively, and $f$ is a constant of order unity\cite{Kratter16}. GI sets in when $Q\leq1$. Eq.\ref{eq:Q} shows that GI is exacerbated in cold and slowly rotating disks with high surface density, or, equivalently when $M_d / M_{\star} > 0.01$\cite{Kratter16}. 

The dominant mode expressed in a GI spiral relates to the degree of instability via the disk to star mass ratio whereby the $m = 4$ dominated pattern, as seen in G358-MM1, occurs at $ M_d/M_{\star} \approx$ 0.25\cite{LodatoRice04}. 
In the treatment of diffusion in density waves by [ref,\cite{LauBertin78}], the relation of $M_d / M_{\star} \approx 1/m$ for massive disks (see also ref\cite{Kratter16}) corroborates a disk to star mass ratio of 0.25 in G358-MM1. Our approach of comparing observed spiral arm characteristics to those of simulated disks is an alternative to directly estimating Q via the measurement of protostar and disk masses, for which optical depth and difficulty isolating emission from the star, disk and core are noted to lead to systematic underestimation of the disk mass in high mass protostars\cite{Forgan18,Jankovic19}.


A simplistic consideration of accretion rates shows that GI is a natural consequence of disk accretion. Infall from the protostellar envelope toward the protostellar disk, $\dot M_{\mathrm{infall}}$, occurs on timescales of the order of free-fall. As material approaches the protostar the conservation of angular momentum counteracts contraction, leading to a transition to slower, disk-mediated accretion where material accretes at a rate of $\dot M_{\mathrm{disk}}$. Infall rates from the envelope onto the disk exceed the steady-state accretion through the disk by approximately an order of magnitude, i.e $\dot M_{\mathrm{infall}} \gg \dot M_{\mathrm{disk}}$, naturally leading to over-dense disks. The growing disk mass leads to higher disk to star mass ratios, thus pushing the system toward GI\cite{Harsono11}. 

Simulations show that rapid angular momentum transport via \emph{low-m} spiral instabilities are the dominant mechanism by which runaway build-up of disk mass is avoided\cite{Harsono11}. Beyond ratios $ M_d/M_{\star} \approx$ 0.25\cite{LodatoRice04}, a series of ``recurrent spiral episodes" act to process material through the massive disk while simultaneously providing shock heating\cite{LodatoRice05,Jankovic19} and reducing the overall surface density\cite{Kratter16}. As can be seen in Eq.1 these processes induce self-regulation which increases Q and restores the disk to a quasi-stable state, preventing fragmentation into proto-stellar companions, a process which occurs at $M_d \approx M_*$\cite{Kratter10}.

Fragmentation divides the available accretable mass into two or more lower mass companions therefore working against the formation of high-mass stars. The physical conditions of high-mass protostellar disks may support against such fragmentation within radial distances of 1200 AU\cite{Klassen16}. Our conclusion is that the disk in G358-MM1 is gravitationally stable enough to avoid runaway disk fragmentation, while being sufficiently GI to provide angular momentum transport via the $m = 4$ mode, i.e. that the star to disk mass ratio is presently in the range of $0.25 < M_d/M_{\star} < 1$.

For such quasi-stability to be maintained it is necessary that infall from the envelope be adequately processed through the disk on long timescales. However, \emph{steady state} continuous disk accretion rates in high-mass protostars are consistently lower than infall rates \cite{Beltran16}. Values of $\dot M_{\mathrm{infall}} = 10^{-4}$ to $10^{-3} M_{\odot}~ \rm yr^{-1}$ [refs,\cite{Beuther17,Sanna19,Motogi17,Motogi19}] have been estimated in observations and values of $\dot M_{\mathrm{infall}} = 2-3 \times 10^{-4} M_{\odot}~ \rm yr^{-1} $ were derived in recent simulations\cite{Meyer17}. Considering the contributions of steady state disk accretion, $\dot M_{\mathrm{steady disk}}$, with the addition of burst accretion, $\dot M_{\mathrm{burst}}$, we can require that, when averaging over a long time, $\dot M_{\mathrm{steady disk}} + \dot M_{\mathrm{burst}} \approx \dot M_{\mathrm{infall}}$, allowing estimates to be made of these parameters. 

The steady state mass accretion rate through the disks is calculated by $ \dot M_{steady disk} = 3 \alpha c_s^3 / (QG) $ [ref,\cite{Frank02}] where $\alpha$ is a viscosity term\cite{ShakuraSunyaev73}. 
Q values of 0.6 and 1 correspond to scenarios of disk fragmentation and spiral arm appearance respectively\cite{Meyer18}. Considering parameters that maximise the accretion rate estimate, $\alpha=0.1$ and $Q=0.6$, and a disk temperature of $T=100$K in agreement with models of HDO and HNCO emission in G358-MM1\cite{Chen20a} and the disk model of [ref\cite{Stecklum21a}], we estimate a steady state disk accretion rate of $\dot M_{steady disk} =$ 2.56 $\times 10^{-5} M_{\odot}~ \rm yr^{-1}$ at a radius of $\sim 1000$ AU. This radius represents a bottleneck where spherical accretion transitions to disk accretion. Even when adopting these parameters, the steady state disk accretion rate remains an order of magnitude below the $10^{-3}$ - 10$^{-4} M_{\odot}~ \rm yr^{-1}$ needed to avoid runaway fragmentation by sufficiently processing envelope infall. Our conclusion is that, in order to avoid disk fragmentation, the spiral density waves in G358-MM1 would need to provide angular momentum dissipation at a rate of an order of magnitude larger than the steady state disk accretion rate. This is consistent with models of the accretion rates in high-mass protostellar disks with spiral pattern instabilities\cite{Harsono11}.

\subsection{GI-induced spirals and accretion bursts}
 
 \null

As a high mass protostar becomes more massive its disk may only achieve the conditions necessary for GI ($M_d / M_* > 10^{-2}$ [ref, \cite{Kratter16}]) by accumulating subsequently higher disk masses. Thus if the accretion rates from the envelope onto the disk remain roughly constant as the stellar mass grows then accretion bursts will become less frequent but more intense over time\cite{Meyer21a}. 
The G358-MM1 accretion burst exhibited a modest mass accretion of 0.566 Jupiter masses\cite{Stecklum21a}, comparable with a small fragment or a spiral arm segment rather than a companion-mass object. The protostellar mass is also low in comparison to other high-mass protostars with observationally confirmed accretion bursts ($\sim 10 M_{\odot}$ for G358-MM1 (this work, see also ref \cite{Chen20b,Stecklum21a}); $\sim 20 M_{\odot}$ for S255IR-NIRS3\cite{Garatti17}; $\sim 20 M_{\odot}$ for NGC 6334I MM1\cite{Hunter17}). The modest burst in G358-MM1, and its comparatively modest protostellar mass are consistent with the expected characteristics of an accretion event occurring in the early phase of high mass protostellar evolution. Indeed, the youth of G358-MM1 has been corroborated by numerous other observational investigations\cite{Stecklum21a,Brogan19a,Bayandina22}.


Our results therefore support the view that disk fragmentation into companion-mass objects likely occurs later in the evolution of high mass protostars when spiral arms become more tightly wound\cite{Cossins09}, exhibit larger amplitudes of \emph{high-m} spiral modes and therefore become less effective at transporting angular momentum\cite{LodatoRice04,LauBertin78}. This gradual formation of larger fragments as the system evolves may explain why larger bursts were seen in the more massive S255IR-NIRS3 and NGC 6334I MM1. Investigations of disk substructure in accretion burst high-mass protostars across the mass spectrum will be needed to test this hypothesis. This situation, however, presents an observational challenge since the gradual shift in dominance from the more readily resolvable \emph{low-m} disk spiral modes to more difficult to resolve \emph{high-m} modes makes detecting spiral arms in high-mass protostars more difficult as they evolve. Synthetic observations show that disks with 4-arm spirals become undetectable with ALMA beyond source distances of 3 kpc\cite{Jankovic19}. The trend for higher-mass protostars to exhibit higher $m$ mode spiral instabilities likely contributes to the lack of detections of spiral arms in high-mass protostellar systems despite the high sensitivity and sub-arcsecond angular resolution of ALMA.


The importance of angular momentum transport is not limited to the outer disk and must occur at all radii of a protostellar system in order for accreting material to reach the central protostar without spinning it up destructively. Results from G358-MM1 imply that the spiral arm density waves in the disks of high-mass protostars can, at least episodically, provide high accretion rates from the outer disk to the inner disk. 
Inward of this region higher ionisation rates cause magneto-hydrodynamic effects to become important, consequently giving rise to disk winds\cite{Hirota17} and rotating jets\cite{Burns15a} which may contribute to angular momentum transfer in the inner disk region. It was recently confirmed that surplus accretion material is launched from this region during an accretion burst, forming a detectable radio jet (S255IR-SMA1 [ref\cite{Cesaroni18}]) which appeared $\sim$1 yr following the accretion burst. Indeed, a flare of water masers, which typically associate with protostellar jets\cite{Moscadelli19}, was seen in G358-MM1 shortly following the methanol maser flare\cite{Bayandina22a}.
To investigate new jet formation and the angular momentum transport mechanisms operating at the inner disk radii new observations capable of detecting disk winds and jet rotation should presently be pursued in G358-MM1.

\subsection{Heat-wave mapping and outlook}
 
 \null

We have shown that heat-wave mapping of 6.7 GHz methanol masers can be used to resolve the spatio-kinematics of accretion disks at the angular resolution necessary to characterise disk substructures in high-mass protostars. This new observational approach compliments ALMA observations of dust and thermal gas and should be attempted again in future accretion burst events. The completion of the SKA and ngVLA will make further progress in the understanding of disk-aided high-mass star formation.
Increasing the number of such high-resolution investigations will be key to refining the next generation of simulations and models of high-mass star formation. Particular focus now needs to extend to the interplay between various mechanisms in which angular momentum transport enables accretion, and the rates at which that accretion occurs, at various disk radii, from the protostellar envelope to the protostellar surface. Ultimately the rates of accretion through each structure in the protostellar system will determine its stability, multiplicity and the final mass of the star.

\begin{figure*}[ht!]
\begin{center}
  \includegraphics[width=0.77\textwidth]{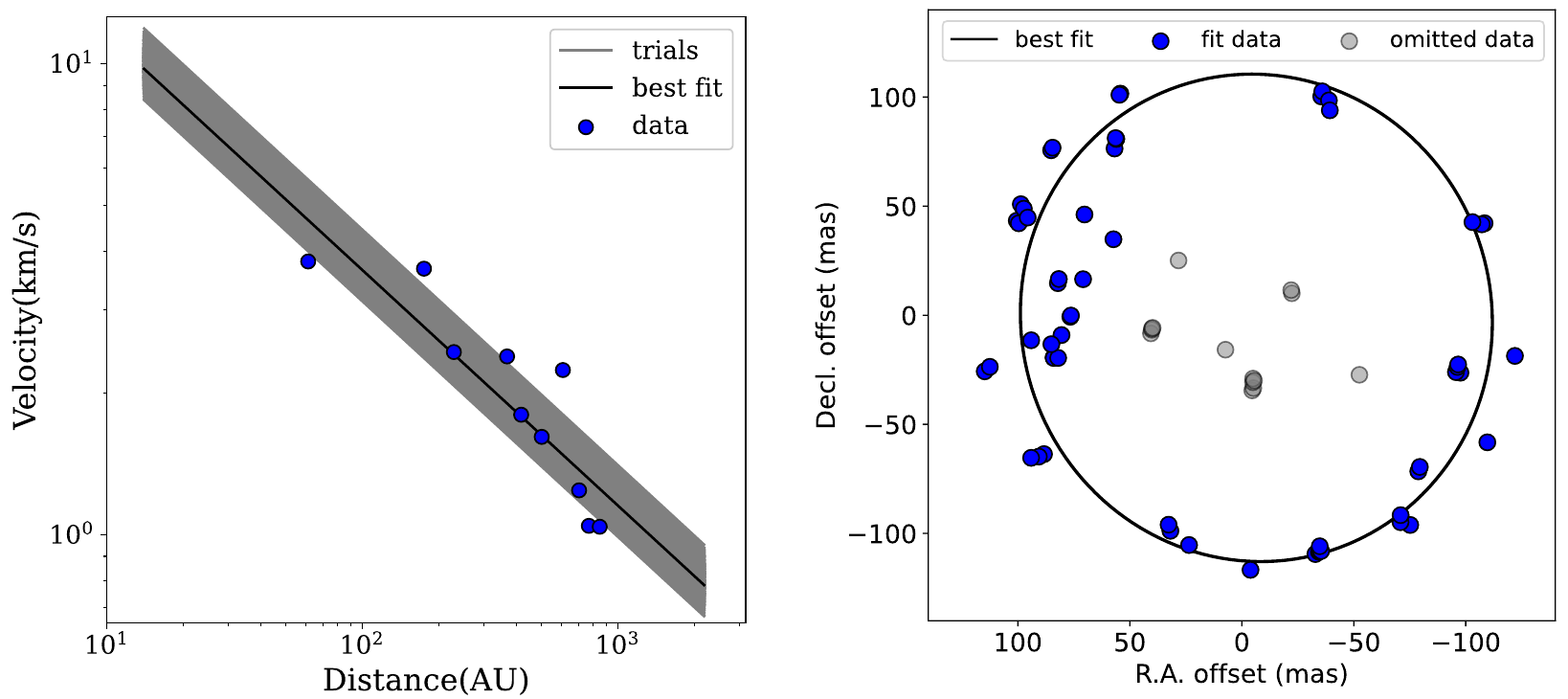} 
\end{center}
\caption{ {\bf Analyses of maser data} (\emph{Left}) Shows the Position-velocity diagram for the maser data an best fit line to the Keplerian profile. Bootrapping trials are shown as semi-opaque lines, indicating the fit uncertainty. \emph{Right} shows the maser spot positions for the outer 50\% of masers in the 5th VLBI epoch (blue circles) where the point size represents the 2.60 milliarcsecond positional uncertainty in the astrometric positions. Spots omitted from the ellipse fitting are shown in grey. The black line shows the ellipse fit to the maser data to determine the system inclination.  \label{Fig3}}
\end{figure*}

\section*{Methods}
~

\noindent \textbf{Observations.} All data used in this work were acquired using very long baseline interferometry (VLBI) pursued in response to the 6.7 GHz methanol maser flare\cite{Sugiyama19} reported to the M2O communications network. Telescope time was sought from multiple VLBI arrays in order to maximise observing cadence in consideration of the limited target of opportunity (ToO) and directors discretionary time (DDT) available at each facility. A summary of the observations used in this work, detailing: facilities, observing dates, and basic observing parameters, is included in Table~\ref{TAB1}.

While observation setups for each array varied somewhat to accommodate differences in observing bandwidth, correlator spectral resolution and observation duration (limited by source elevation at the site of the VLBI array), all observations consisted of alternating scans between the target G358-MM1 ($\alpha , \delta) =$ (17:43:10.0000, -29:51:45.800) and a phase reference continuum source J1744-3116  ($\alpha , \delta$) = (17:44:23.5782, -31:16:36.292) in phase referencing cycles of 4.5 minutes, recording in both left and right hand circular polarisation. Data were correlated in two passes, once utilising the full available bandwidth of the facility at coarse spectral resolution in order to maximise the detection of continuum sources, and again using `zoom band' filtering of a 16 MHz spectral window centered at the methanol ($5_1 \xrightarrow{} 6_0~A^+$) maser emission frequency of 6.668 GHz. In all data sets the continuum sources were sufficiently detected using only the zoom band data; wide-band coarse resolution correlation data were not necessary and therefore not used but are publicly available.

\null 

\noindent \textbf{Data reduction.} All data were processed using the Astronomical Imaging Processing Software (AIPS). While variations exist in the practices of processing data from each of the facilities used in this work, the general methodology is essentially identical. Data were loaded into AIPS and erroneous data were flagged based on observation logs provided by facilities and observers in addition to channels and time periods containing spurious data. A-priori telescope gain calibration was performed using Tsys and gain curve information provided by stations, either in the form of `antab' files (EVN, LBA) or utilising the TY and GC tables that were shipped with the data (VLBA). Stations for which no such calibration information was retrievable were gain-scaled to match well calibrated stations via comparison of the maser autocorrelation spectra using the AIPS task ACFIT.

Absolute determination of the relative delay in signals arriving at each station was performed via the AIPS task FRING using a bright calibrator source observed at roughly hourly intervals. At this stage slowly varying delay residuals such as VLBI clock drifts were also calibrated, and the instrumental phase-delay difference between each polarisation was also solved. Rate solutions acquired at this stage were discarded. Phase referencing was carried out by calibrating residual phase and rate solutions on the referenced source J1744-3116 and applying the aforementioned delay, phase and rate solutions to the maser data. The IRFC coordinates of the maser were then determined in initial imaging with the AIPS task IMAGR, and further refinement to the phase and rate residual calibration was carried out using self-calibration of a reference maser channel, either using the task CALIB in AIPS if the structure was simple, or using the \emph{selfcal} procedure in the \emph{DifMap} software package for epochs where the reference maser channel indicated non-pointlike structure.

Small relative frequency shifts caused by a difference in the antenna Earth locations during rotation and the Earth-centered reference point used in the correlator were corrected using the AIPS task CVEL. Using the AIPS task IMAGR, image cubes in the RA, DEC and LSR velocity coordinates were produced in the ranges of 1 by 1 arcseconds in the spatial domain and $\pm 50$ km s$^{-1}$ around the brightest maser channel in the spectral domain, using natural weighting to accommodate resolved emission. Moment maps were produced for each epoch using AIPS task MOMNT and are shown in Figure~\ref{Fig1}. Image cubes were then searched for maser emission via an automated 2D Gaussian fitting procedure in AIPS called SAD, to ascertain the coordinates and flux densities, along with value uncertainties, for maser emitting regions in each frequency channel. Gaussian fitting was applied to emission above a 7 $\sigma$ detection limit which avoided sidelobe inclusion into the final data set. The product of this procedure, applied to each VLBI epoch, was six machine-readable lists, known as maser `spot maps', that are convenient for plotting and statistical analyses.  Calibrated data sets of the 6.7 GHz methanol maser data in FITS format, each epoch typically a few tens of Mb, and spot map data, are publicly available (see Data Availability).

In order to allow appropriate comparison of the multi-epoch results, an error term accounting for systematic tropospheric and inonospheric phase residuals in each epoch, based on the estimation method described in [ref\cite{Asaki07}], and the positional uncertainty of the phase reference source based on 2D Gaussian fits to their VLBI images, were added in quadrature to the individual maser positional uncertainties.

\begin{figure*}[ht!]
\begin{center}
\includegraphics[trim={0cm 0cm 0cm 0cm},
clip,width=1.0\textwidth]{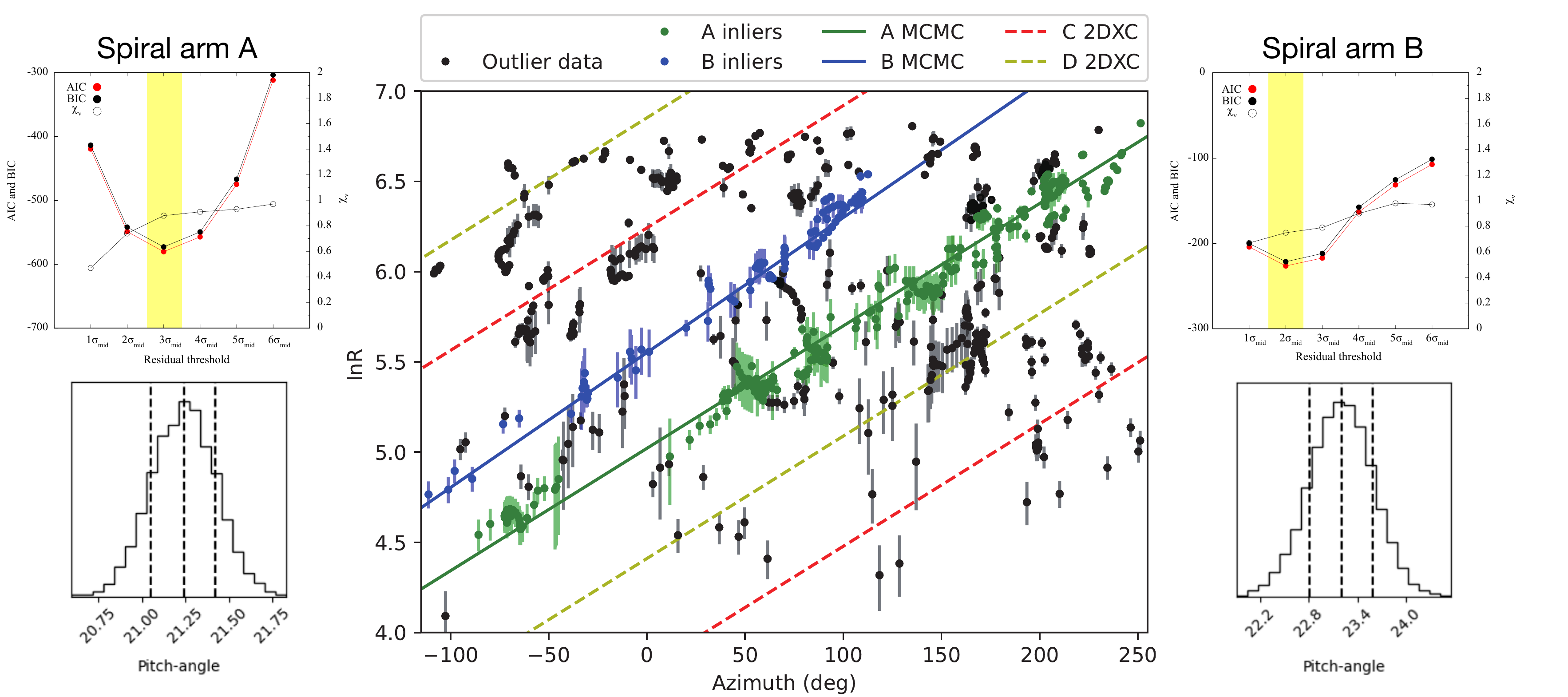}
\end{center}
\caption{ {\bf Identification of spiral arms A and B} (\emph{Upper left}) shows the AIC, BIC and $\chi$ values for 200,000 trials of each residual threshold in the RANSAC procedure for arm A. (\emph{Lower left}) shows the posterior distribution for pitch angles fit during 10,000 samples in the MCMC procedure for arm A. (\emph{Upper and lower right}) show the same for arm B. (\emph{Center}) shows the six combined spotmap data set in $\phi-ln(R)$ space where spots determined by RANSAC to be associated with arms A and B, and best fit arm functions determined by MCMC, are shown in green and blue, respectively. The nomial locations of spiral arms C and D are shown as dashed lines. Colours of arms are consistent with Figure~\ref{Fig5}. Error bars express each maser's astrometric uncertainty.\label{Fig4}}
\end{figure*}

\null 

\noindent \textbf{Parameterizing the Keplerian disk and enclosed mass.} Following the heat-wave mapping technique, combining the individual spotmaps from all VLBI epochs produced a discretely sampled map of the spatio-kinematics of the protostellar disk. The largest velocity gradient in the combined spotmap was at a position angle of $45^{\circ}$ from equatorial north, in the right ascension direction, which was adopted as the perpendicular to the rotation axis of motion. To determine the radial rotation profile of the maser-traced structure its position-velocity data were taken from a cut perpendicular to the rotation axis (see Figure~\ref{Fig2}, \emph{middle}). 
Blueshifted and redshifted maser velocities, relative to the systemic velocity of -16.5 km s$^{-1}$, were of similar magnitude, validating the source velocity determined by [ref\cite{Brogan19a}]. The modulus of each maser’s velocity relative to the systemic velocity was taken, effectively folding velocity data to the positive velocity offset direction. Masers inside $R<50$ AU, which are not well sampled by heat-wave mapping, and which are near the solid-body rotation portion of the Keplerian velocity profile, were discarded. Maser position-velocity data in the 50 to 1000 AU range were binned on the position axis at widths of 88 AU where the bin width was decided in such a manner to maximise the number of bins while ensuring that all position bins had at least one velocity data point. Velocity data were already quantised at the spectral resolution at which VLBI data were correlated (See Table 1) and were not binned further. The maximum velocity in each position bin was taken to determine the outer envelope of the velocity profile - corresponding to the position-velocity envelope at which the rotation curve is evaluated.

A power law was found for the position-velocity data via a least-squares fit, simultaneously solving for the slope and intercept of a straight line through the data in position-log($v$) space. Bootstrapping analyses (See Figure~\ref{Fig3}, \emph{Left}), employing 500 trials, determined a velocity profile index of $v \propto r^{-0.510 \pm 0.095}$ and an intercept of 1.28$\pm$3.60, respectively, confirming with high confidence that the 6.7 GHz masers trace Keplerian rotation. 
Data were then re-analysed with a fixed Keplerian power law ($v \propto r^{-0.5}$) to evaluate the enclosed mass of the system with higher precision, which was determined by a bootstrapped least-squares fit to be $M \times \sin{i}^2 = 1.53 \pm$0.02 $M_{\odot}$, where $M$ is the dynamic mass and $i$ is the system inclination. The Keplerian profile produced by these parameters is plotted over the original data in the position-velocity domain in Figure~\ref{Fig2} (\emph{Right}).

In order to remove the dependency of the measured enclosed mass on the system geometry an ellipse was fit to the spotmap of the 5th VLBI epoch, which shows the most extended and complete ring distribution, to determine the disk inclination using linear regression. An ellipse fit employing 10,000 bootstrapping trials was conducted on the spotmap, leading to an inclination estimate of $i = 26.8 \pm 8.6 ^{\circ}$ (with $i=0^{\circ}$ representing a face-on disk). Repeating the fit and bootstrapping, while omitting the inner 50\% of masers in order to isolate the outer radius of the disk, gave a consistent but more precise measurement of the disk inclination of $i = 21.4 \pm 4.7 ^{\circ}$. The latter value, derived using only masers that trace the ring distribution (see Figure~\ref{Fig3}, \emph{right}), was adopted.
Subtracting inclination effects from the enclosed mass derived from fitting the velocity profile of the maser data lead to an enclosed mass estimate of $M = 11.5 \pm 4.8$ M$_{\odot}$. Both the inclination and enclosed mass determined in this work are consistent with those estimated from VLA data\cite{Chen20b} and interpretation of the SED\cite{Stecklum21a}.

\noindent \textbf{Modelling the spiral arm disk substructure.} The six epoch combined maser spot map data set was analysed in $\phi-ln(R)$ space, where $R$ is the radial distance of the maser spot from the map center and $\phi$ is the azimuth angle of the maser spot counted clockwise from the West direction. In $\phi-ln(R)$ space, spiral arms are effectively `unwound' and appear as straight lines (see Figure~\ref{Fig4}, \emph{middle}), enabling conventional fitting routines as is commonly used for characterising spiral arm patterns in disks\cite{Lee20}. One spiral arm clearly dominated the $\phi-ln(R)$ space (denoted as arm A, shown by green data points in Figure~\ref{Fig4}, \emph{middle}).

\begin{figure*}[ht!]
\begin{center}
\hspace{-0.5cm}\includegraphics[width=0.90\textwidth]{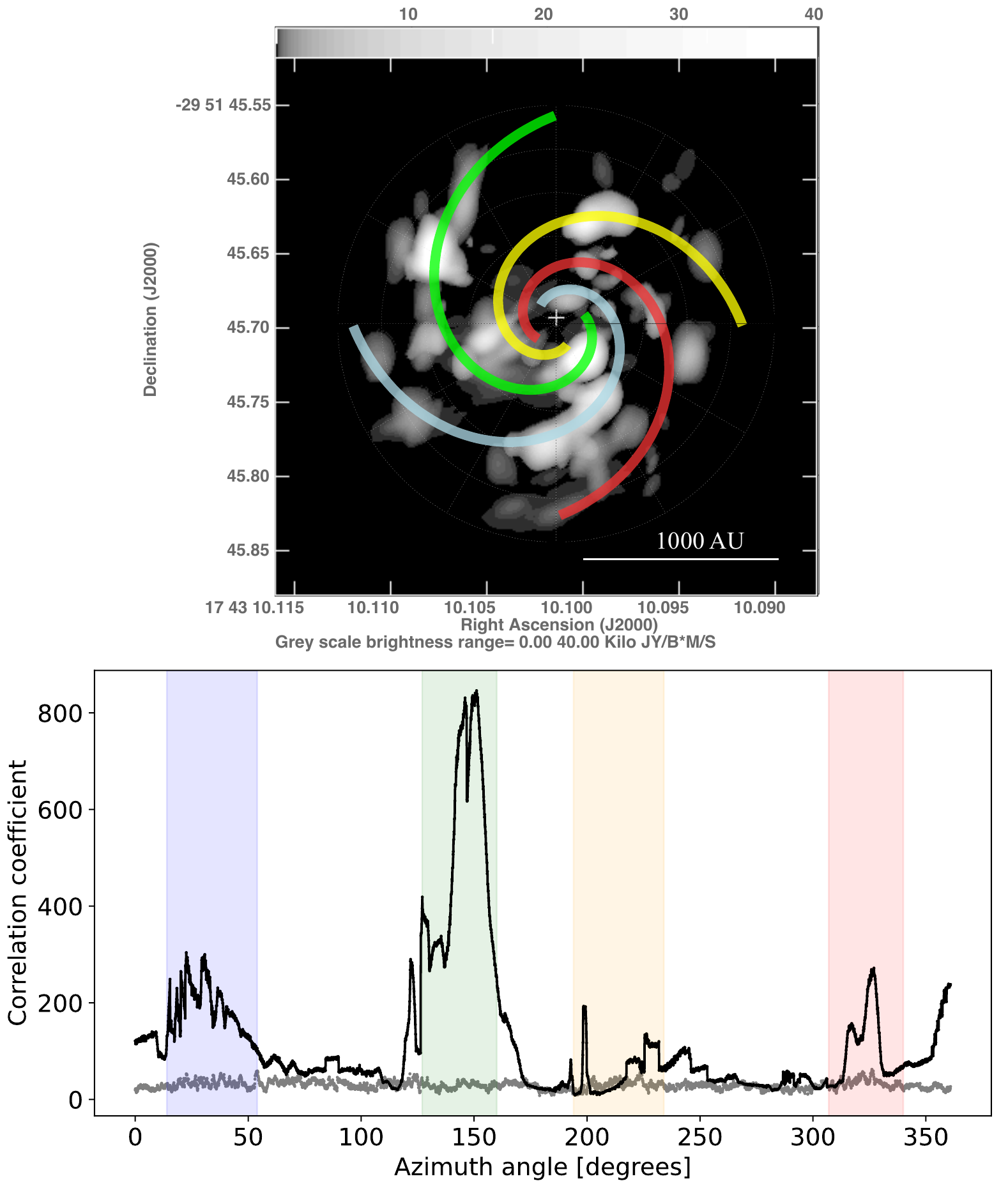}
\caption{ {\bf 4-arm spiral identification} \emph{Below} shows the results of the spatial 2D cross-correlation as a function of azimuth angle. The green and blue regions indicate the full-width half-maximum range for the correlation peaks associated with arms A and B, respectively. Red and yellow regions highlight the disk regions at $\pm 180^{\circ}$ opposite to the green and blue regions where symmetric arm pairs were searched. The grey line shows a $5\sigma$ detection criteria derived from the 5 times the standard deviation of correlation coefficients at each azimuth acquired from 10 sets of random data of the same size and variable ranges as the maser data. (\emph{Above}) shows the spiral structure model in G358-MM1 plotted on the flux density map (grey-scale). Green and blue arms represent arms A and B respectively, parameterised by RANSAC and MCMC. The red and yellow lines illustrate arms C and D, which represent the symmetric pairs of arms A and B, respectively, detected using 2D cross-correlation. \label{Fig5}}
\end{center}
\end{figure*}

Data corresponding to arm A were identified using the Random Sample Consensus (RANSAC) algorithm of the Python library \emph{scikit-learn}, conducting 200,000 trial fits using \emph{LineModelND} and setting an initial residual threshold of $1.5 \sigma$. The procedure defined two subsets; the inliers: data points which identify with the given test model parameter set, and outliers. The inlier data for arm A were then fit with a straight line model using a Monte Carlo Markov Chain (MCMC) algorithm from Python's \emph{emcee} package, employing 10,000 samples. The spiral arm function was defined as ln(R) = ln(R$_{\mathrm{ref}}$) + ($\phi$ - $\phi_{\mathrm{ref}}$) * tan (Pitch-angle), where $R_{\mathrm{ref}}$ is the radius where the arm crosses $\phi=\phi_{\mathrm{ref}}$ and the pitch angle is a straight line in $\phi-ln(R)$ space.

Exploring the parameter space of the residual threshold parameter (1,2,...,6 $\sigma$) set in the RANSAC stage, the best value was evaluated based on the Akaike Information Criterion (AIC), Bayesian information criterion (BIC) and reduced $\chi^2$ of the MCMC fit for each iteration. The effect of the threshold is shown in Figure~\ref{Fig4} (\emph{upper left}) where the best fit for arm A was achieved for a 3$\sigma$ residual threshold. Under this condition the MCMC procedure identified the pitch angle of arm A as $ \theta = 21\fdg 2\pm0\fdg2$ where the error is determined by $\chi^2$ normalisation. Figure~\ref{Fig4} (\emph{lower left}) shows the parameter space exploration of the pitch angle, evaluated by the likelihood function in the MCMC fitting.

The data points associated with arm A, which had been noted as inliers during the RANSAC stage, were then removed from the original data set and the whole procedure was repeated to reveal a second spiral arm, arm B, which had a pitch angle of $\theta = 23\fdg2\pm0\fdg4$ and trails the first arm by $\sim 90 ^{\circ}$. Figure~\ref{Fig4} (\emph{upper and lower right}) show the RANSAC residual threshold best value and the likelihood function as a function of pitch angle, respectively, for arm B.

\begin{figure*}[ht!]
\vspace{-0.77cm}
\begin{center}
\includegraphics[width=1.0\textwidth]{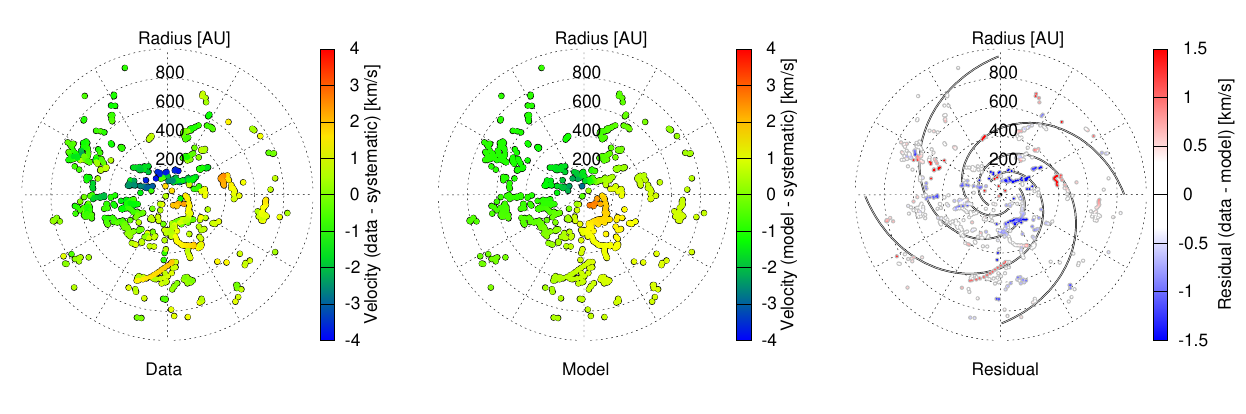}
\end{center}
\caption{ {\bf Velocity residuals upon subtracting the Keplerian disk model} \emph{Left} shows the original astrometric data, \emph{Middle} shows the expected velocities at the locations of detected masers, assuming purely Keplerian rotation. \emph{Right} shows the residual velocities after subtracting the velocities of the model from the data. The expected signature of an interaction-induced warp is absent in the residual map. The spiral arm pattern discussed in this work is superposed on the residual map, indicating no obvious association of super- nor sub-Keplerian motions with the arms. \label{Fig6}}
\end{figure*}

The pitch angles of logarithmic spiral arms A and B were similar, as is expected for GI induced structure, but their azimuthal locations were near a $\pi/2$ radian separation rather than the symmetric configuration expected for GI. In order to uncover any additional arm pairs the data were again searched using a more sensitive 2D spatial cross-correlation procedure, \emph{correlate2d}, from the SciPy package.
This simple cross-correlation procedure cannot determine spiral arm shape parameters but is highly effective at identifying matches to known patterns (i.e. known spiral arm functions) in low signal to noise data. Arm A parameters were adopted as a model prior expressed as a function, 
$lnR = (-0.00678\pm0.00007)\phi+(5.020\pm0.008)$, sampled in Cartesian coordinates with 10,000 points. This prior was used for correlating with the full spot map data. 
Preceding correlation the data were binned in Cartesian coordinates to a uniform $30 \times 30$ cell grid in preparation for the discrete Fourier transform used in cross-correlation. Correlation was then performed in Cartesian coordinates for rotation angles between 0$^{\circ}$ and 360$^{\circ}$ in steps of 0.1$^{\circ}$, giving the correlation coefficient as a function of azimuth angle, which is shown as a black line in Figure~\ref{Fig5} (\emph{below}), in which several peaks can be seen. To evaluate the significance of the peaks the process was repeated using 10 sets of randomly generated data of azimuth and radii in the same range ($0 < \rm{azimuth} <360 ^{\circ}$; $0 < R < 918$ AU) and equivalent number of data points as the maser data. The 2D cross-correlation process was then repeated following the same procedure outlined above. The null rejection level, defined here as 5 times of the standard deviation of correlation coefficients at each azimuth value for the random data sets, is shown as a grey line in Figure~\ref{Fig5} (\emph{below}).

Arms A and B, which were identified by RANSAC and parameterised by MCMC, appear in Figure~\ref{Fig5} (\emph{below}) as prominent correlation peaks and are highlighted by green and blue regions, respectively, where azimuth ranges were determined as the full-width half maximum of the peaks. Yellow and red regions represent the symmetrically opposite ($\pm 180^{\circ}$) disk regions where symmetric pairs to arms A and B may be expected in the case of a rotationally symmetric spiral pattern. The analysis revealed prominent peaks corresponding to arm detections at the opposite locations to arms A and B, which which are designated as arms C and D, respectively. 

Figure~\ref{Fig5} (\emph{above}) shows the flux density map of 6.7 GHz methanol maser emission in G358-MM1. Superimposed are arms A, B, C and D plotted in green, blue, red and yellow, respectively. Arm morphologies for arms A and B were defined using parameters determined by the RANSAC and MCMC procedures described above. Arms C and D, identified by the 2D correlation, are plotted in red and yellow respectively, adopting the same morphology as arms A and B with the addition of a $\pi$ radians rotation. The majority of maser emission identifies with one of the four spiral arms, however some inter-arm maser emission is also apparent, possibly indicating the existence of minor arm spurs or emission from masers above or below the disk-plane.

\null

\subsection{Data availability}
Data used in this work can be accessed by searching experiment codes (See Table 1) in the following data archives:\\
LBA data: (https://atoa.atnf.csiro.au)\\
EVN data: (http://archive.jive.nl/scripts/portal.php)\\
VLBA data: (https://https://data.nrao.edu/portal/\#/)

\noindent The maser spot maps used in this work, in addition to calibrated data from the six epochs in FITS format, are available on the following link: \\ https://www.masermonitoring.com/g358-mm1-data-availability

\null 

\subsection{Code availability} The \emph{correlate2d} algorithm is available from Python's  \emph{SciPy} package. The Monte Carlo Markov Chain (MCMC) algorithm is available from Python's \emph{emcee} package. The Random Sample Consensus (RANSAC) algorithm is available from the Python's \emph{scikit-learn} package.

\addtocounter{enumiv}{\value{firstbib}}

\null 

\subsection{Acknowledgements} - 

R.A.B. acknowledges support through the EACOA Fellowship from the East Asian Core Observatories Association.
R.A.B., J.O.C. and G.C.M. acknowledge the Global Emerging Radio Astronomy Foundation (GERAF) for contributions to radio astronomy.
T.Hirota is financially supported by the MEXT/JSPS KAKENHI Grant Numbers 17K05398, 18H05222, and 20H05845.
Y.Y. is financially supported by the MEXT/JSPS KAKANHI Grant Numbers 21H01120 and 21H00032.
L.U. acknowledges support from the University of Guanajuato (Mexico) grant ID CIIC 164/2022. 
A.C.G. acknowledges support by PRIN-INAF-MAIN-STREAM 2017.
M.O. thanks the Ministry of Education and Science of the Republic of Poland for support and granting funds for the Polish contribution to the International LOFAR Telescope (arrangement no. 2021/WK/02) and for maintenance of the LOFAR PL-612 Baldy (MSHE decision no. 28/530020/SPUB/SP/2022).
A.B., M.D. acknowledge support from the National Science Centre, Poland through grant 2021/43/B/ST9/02008
O.B. acknowledges financial support from the Italian Ministry of University and Research - Project Proposal CIR01\_00010.
A.M.S. and D.A.L. were supported by the Ministry of Science and Higher Education of the Russian Federation (state contract FEUZ-2023-0019).

\newpage 

\subsection{Author contributions} - \\ R.A.B. lead the project as principle investigator for the observations, processed the VLBI data, and authored the manuscript. Y.U performed the Keplerian modelling of the maser data. N.S. performed the spiral arm identification procedures using RANSAC and MCMC. J.Blanchard conducted the 2D cross-correlation for identification of additional spiral arms. Z.F. conducted the disk inclination measurement. K.S. and Y.Y. selected the target maser source. E.V., J.B, F.v.d.H, Y.Y., Y.T., A.A., G.C.M., M.O., M.D., conducted single-dish monitoring of masers toward G358.93-0.03. G.O., S.P.E., L.H. and C.P. conducted the LBA observations. All authors contributed to the scientific discussion and helped with the authorship and reviewing process of the manuscript.

\noindent \textbf{Additional information}
{\bf~\\Reprints and permissions information} is available at \url{www.nature.com/reprints}.
{\bf~\\Correspondence and requests for materials} should be address to R. A. Burns (ross.burns@nao.ac.jp).

\noindent \textbf{Competing interests} The authors declare no competing financial interests.

\null 

\noindent This manuscript was prepared using the author's custom LaTeX template. Research contents reflect the state of the paper before proofs. The final article published by Springer Nature should be taken as the most complete version.

\onecolumn

\newpage

\section{Author affiliations}

\begin{affiliations} - \\
 \item Mizusawa VLBI Observatory, National Astronomical Observatory of Japan, 2-21-1 Osawa, Mitaka, Tokyo 181-8588, Japan \\
 \item Department of Science, National Astronomical Observatory of Japan, 2-21-1 Osawa, Mitaka, Tokyo 181-8588, Japan \\
\item Korea Astronomy and Space Science Institute, 776 Daedeokdae-ro, Yuseong-gu, Daejeon 34055, Republic of Korea \\
\item Department of Physics, National Chung Hsing University, No. 145, Xingda Rd., South Dist., Taichung 40227, Taiwan \\
\item National Astronomical Research Institute of Thailand (Public Organization), 260 Moo 4, T. Donkaew, A. Maerim, Chiangmai 50180, Thailand \\
\item NRAO, PO Box O, 1003 Lopezville Rd., Socorro, NM 87801, USA \\
\item Department of Physics, Faculty of Science, University of Malaya, 50603, Kuala Lumpur, Malaysia \\
\item Joint Institute for VLBI ERIC, Oude Hoogeveensedijk 4, 7991 PD Dwingeloo, The Netherlands \\
\item Center for Astronomy, Ibaraki University, 2-1-1 Bunkyo, Mito, Ibaraki 310-8512, Japan \\
\item Department of Astronomical Sciences, SOKENDAI (The Graduate University for Advanced Studies), 2-21-1 Osawa, Mitaka-shi, Tokyo 181-8588, Japan \\
\item University of Science and Technology, Korea (UST), 217 Gajeong-ro, Yuseong-gu, Daejeon 34113, Republic of Korea \\
\item Ventspils International Radio Astronomy Center", Ventspils University of Applied Sciences, Inzenieru Str. 101, Ventspils, LV-3601, Latvia \\
\item Radio Astronomy and Geodynamics Department of Crimean Astrophysical Observatory, Katsively, RT-22 Crimea Ukraine \\
\item Institute of Astronomy, Faculty of Physics, Astronomy and Informatics, Nicolaus Copernicus University, Grudziadzka 5, 87-100 Torun, Poland \\
\item INAF-Osservatorio Astronomico di Capodimonte Napoli, Salita Moiariello 16, 80131 - Naples, Italy \\
\item Astronomical Observatory, Ural Federal University, 19 Mira Street, 620002 Ekaterinburg, Russia \\
\item Th\"uringer Landessternwarte, Sternwarte 5, 07778 Tautenburg, Germany \\
\item NRAO, 520 Edgemont Rd, Charlottesville, VA, 22903, USA \\
\item Australia Telescope National Facility, CSIRO, PO Box 76, Epping NSW 1710, Australia \\
\item NRC Herzberg Astronomy and Astrophysics, 5071 West Saanich Rd, Victoria, BC, V9E 2E7, Canada \\
\item INAF Osservatorio Astronomico di Cagliari, Via della Scienza 5, 09047 Selargius, Italy \\
\item Department of Physical Sciences, The Open University of Tanzania, P.O. Box 23409, Dar-Es-Salaam, Tanzania \\
\item Hartebeesthoek Radio Astronomy Observatory, PO Box 443, Krugersdorp, 1741, South Africa \\
\item Max Planck Institute for Astronomy, K\"onigstuhl 17, 69117 Heidelberg, Germany \\
\item Space Research Unit, Physics Department, North West University, Potchefstroom 2520, South Africa \\
\item Department of Physics and Astronomy, Faculty of Physical Sciences, University of Nigeria, Carver Building, 1 University Road, Nsukka, Nigeria \\
\item INAF - Istituto di Radioastronomia \& Italian ALMA Regional Centre, Via P. Gobetti 101, 40129 Bologna, Italy \\
\item School of Natural Sciences, University of Tasmania, Private Bag 37, Hobart, Tasmania 7001, Australia \\
\item Departamento de Astronom\'ia, Universidad de Guanajuato, A.P. 144, 36000 Guanajuato, Gto., Mexico \\
\item Space Radio-Diagnostic Research Center, Faculty of Geoengineering, University of Warmia and Mazury Oczapowskiego 2, PL-10-719 Olsztyn, Poland \\
\item INAF – Osservatorio Astrofisico di Arcetri, Largo E. Fermi 5, 50125 Firenze, Italy \\
\item SKA Observatory, Jodrell Bank, SK11 9FT, UK \\
\item Center for Astrophysics, GuangZhou University, Guangzhou, China \\
\item Shanghai Astronomical Observatory, Chinese Academy of Sciences, Shanghai, China \\
\end{affiliations}


\end{document}